\documentclass[3p,times,twocolumn]{elsarticle}

\usepackage{ecrc}

\volume{00}

\firstpage{1}

\journalname{Nuclear Physics B Proceedings Supplement}

\runauth{A. Vicente}

\jid{nuphbp}

\jnltitlelogo{Nuclear Physics B Proceedings Supplement}

\usepackage{amsmath,amssymb}
\usepackage{amsthm}
\usepackage{xspace}
\usepackage{listings}

\usepackage[figuresright]{rotating}

\newcommand\SARAH{{\tt SARAH}\xspace}
\newcommand\PreSARAH{{\tt PreSARAH}\xspace}
\newcommand\FeynArts{{\tt FeynArts}\xspace}
\newcommand\FormCalc{{\tt FormCalc}\xspace}
\newcommand\SPheno{{\tt SPheno}\xspace}

\newcommand\FlavorKit{{\tt FlavorKit}\xspace}
\newcommand\Mathematica{{\tt Mathematica}\xspace}
\newcommand{\Fortran}{\texttt{Fortran}\xspace}
\def\hc{\text{h.c.}}

\begin{document}

\begin{frontmatter}

\dochead{}

\title{FlavorKit: a brief overview}

\author{A. Vicente}
\ead{avelino.vicente@ulg.ac.be}

\address{IFPA, Dep. AGO, Universit\'e de Li\`ege, Bat B5, Sart-Tilman
  B-4000 Li\`ege 1, Belgium}

\begin{abstract}
We give a brief overview of \FlavorKit, a kit for the study of flavor
observables beyond the standard model. In contrast to previous flavor
codes, \FlavorKit is not restricted to a single model, but can be used
to obtain predictions for flavor observables in a wide range of
models.  \FlavorKit can be used in two different ways. The basic usage
of \FlavorKit allows for the computation of a large number of lepton
and quark flavor observables, using generic analytical expressions for
the Wilson coefficients of the relevant operators. The setup is based
on the public codes \SARAH and \SPheno, and thus allows for the
analytical and numerical computation of the observables in the model
defined by the user. If necessary, the user can also go beyond the
basic usage and define his own operators and/or observables. For this
purpose, a \Mathematica package called \PreSARAH has been
developed. This tool uses \FeynArts and \FormCalc to compute generic
expressions for the required Wilson coefficients at the tree- and
1-loop levels. Similarly, the user can easily implement new
observables. With all these tools properly combined, the user can
obtain analytical and numerical results for the observables of his
interest in the model of his choice.
\end{abstract}


\end{frontmatter}

\section{Introduction}
\label{sec:intro}

The recent discovery of the Higgs boson at the LHC
\cite{Aad:2012tfa,Chatrchyan:2012ufa} and the measurement of its
properties and couplings to fermions and gauge bosons
\cite{CMS:2014ega} have led to a global picture in very good agreement
with the Standard Model (SM) predictions. Although there is still room
for deviations from the SM expectations, these are now quite
constrained.

At this point, we should not forget that, besides theoretical
considerations such as the hierarchy problem, there are good
phenomenological reasons to go beyond the SM (BSM). Among others,
these include dark matter, non-zero neutrino masses and mixings and
the baryon asymmetry of the universe. These fundamental problems of
current particle physics and cosmology do not have an explanation
within the SM, and thus motivate the search for new physics.

The unknown origin of the flavor sector of the SM is another good
reason to consider extended BSM scenarios. This program has been
extensively followed in the quark sector and, more recently, in the
lepton sector after the discovery of neutrino flavor oscillations. Low
energy experiments focused on flavor observables can play a major role
in the search of new physics, since new particles leave their traces
via quantum effects in flavor violating processes such as $b\to s
\gamma$, $B_s \to \mu^+ \mu^-$ or $\mu\to e \gamma$. Therefore, low
energy experiments devoted to observables from the Kaon- and B-meson
sectors, rare lepton decays and electric dipole moments turn out to be
complementary to the direct searches at high energy colliders, opening
a window to new physics in the high intensity frontier.

Here we present a general overview of \FlavorKit \cite{Porod:2014xia},
a computer tool for the computation of flavor observables. There are
already several public tools for this purpose:
{\tt SuperIso} \cite{Mahmoudi:2009zz},
{\tt SUSY\_Flavor} \cite{Crivellin:2012jv},
{\tt NMSSM-Tools} \cite{Ellwanger:2006rn},
{\tt MicrOmegas} \cite{Belanger:2013oya}, 
{\tt SuperBSG} \cite{Degrassi:2007kj},
{\tt SuperLFV} \cite{Murakami:2013rca},
{\tt SuseFlav} \cite{Chowdhury:2011zr},
{\tt IsaJet} with {\tt IsaTools} \cite{Paige:2003mg,Baer:2003xc} 
and {\tt SPheno} \cite{Porod:2003um,Porod:2011nf}. 
However, these codes are valid only in specific models. In addition,
they can hardly be extended by the user to calculate additional
observables. In contrast, \FlavorKit is not restricted to a single
model, but can be used to obtain predictions for flavor observables in
a wide range of models. This makes \FlavorKit the perfect tool for the
study of flavor physics beyond the SM.

The rest of the manuscript is structured as follows: in the next
section we give a brief review of the calculation of flavor
observables and introduce the terminology used in the rest of the
paper. In Sec. \ref{sec:flavorkit} we present \FlavorKit and explain
how to use it, both for basic and advanced users. Finally, we
summarize in Sec. \ref{sec:summary}.

\section{Computation of flavor observables}
\label{sec:flavorobs}

Once we have chosen a BSM model, the strategy for the computation of
a flavor observable follows three steps:

\begin{itemize}

\item {\bf Step 1:} First, one considers an effective Lagrangian that
  includes the operators that contribute to the flavor observable of
  our interest,
 \begin{equation}
  {\cal L}_{eff} = \sum_i C_i {\cal O}_i \, .
 \end{equation}
 This Lagrangian consists of a list of operators, ${\cal O}_i$, with
 Wilson coefficients $C_i$. These can be induced either at tree or at
 higher loop levels and include both the SM and the BSM contributions
 ($C_i = C_i^\text{SM} + C_i^\text{BSM}$).

\item {\bf Step 2:} The Wilson coefficients are computed
  taking into account all possible tree-level and
  1-loop topologies leading to the ${\cal O}_i$ operators.

\item {\bf Step 3:} The results for the Wilson coefficients are
  plugged in a general expression for the observable and a final
  result is obtained.

\end{itemize}

This general strategy can be easily shown with an example. Let us
consider the computation of BR($\mu \to e \gamma$) in the Standard
Model extended by right-handed neutrinos and Dirac neutrino
masses. The first step is, as explained above, to choose the relevant
operators. In this case, at leading order the radiative decays
$\ell_\alpha \to \ell_\beta \gamma$ only receive contributions from
dipole operators. Therefore, the relevant effective Lagrangian is
given by
\begin{equation} 
{\cal L}_{\mu e \gamma}^{\text{dipole}} = i e \, m_{\mu} \, \bar e \sigma^{\mu \nu} q_\nu \left(K_2^L P_L + K_2^R P_R \right) \mu \, A_\mu + \hc
\end{equation}
Here $e$ is the electric charge, $q$ the photon momentum and $P_{L,R}
= \frac{1}{2} (1 \mp \gamma_5)$ are the usual chirality
projectors. This concludes step 1.

\begin{figure}[tb]
\begin{center}
\includegraphics[width=0.9\linewidth]{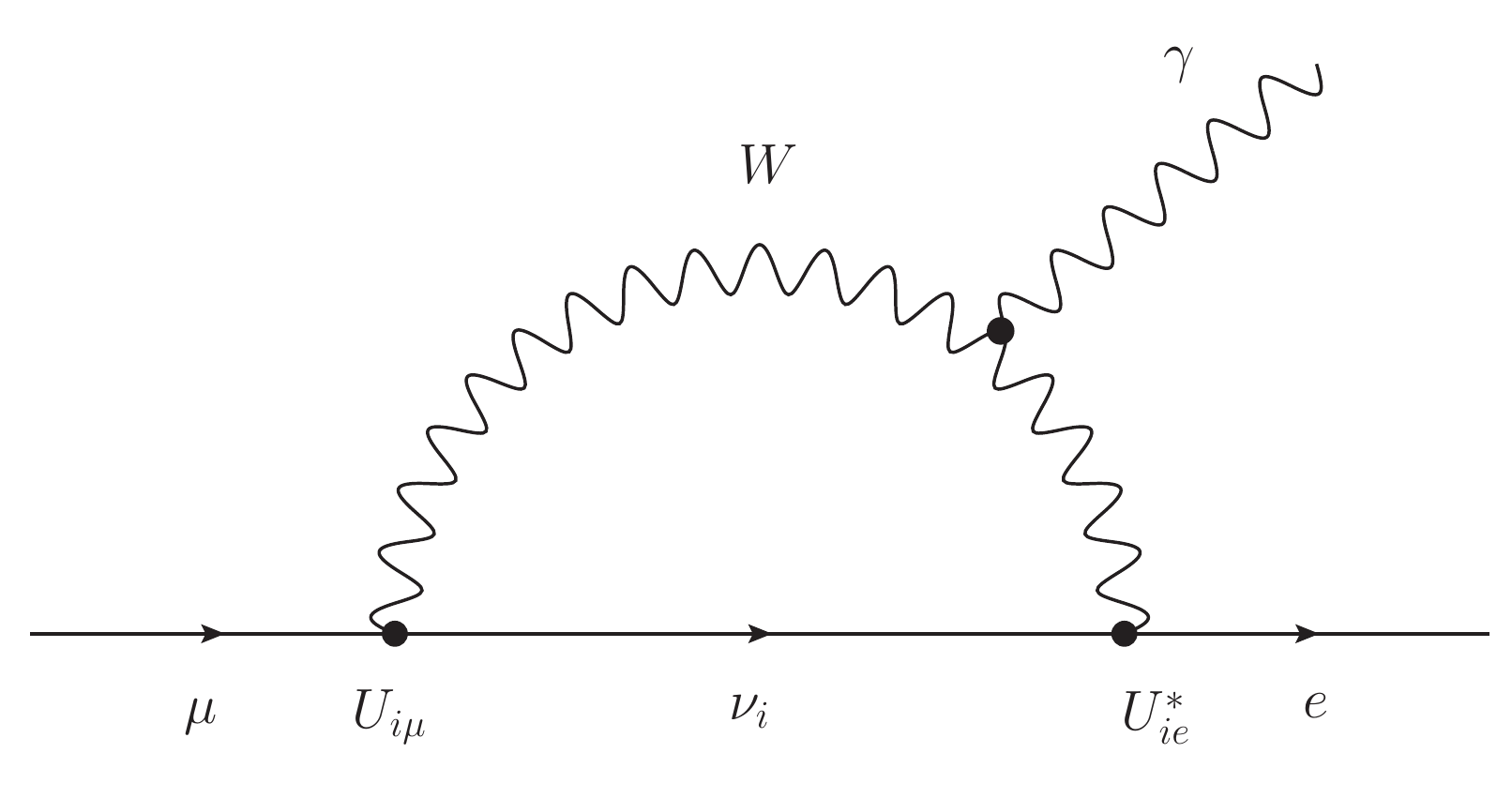}
\end{center}
\caption{1-loop Feynman diagram that contributes to BR($\mu \to e
  \gamma$) in the SM extended with right-handed neutrinos. In addition
  to this diagram, one also has to consider 1-loop Feynman diagrams
  with the photon line attached to the external leptons.}
\label{fig:nuloop}
\end{figure}

In the next step, one must compute the Wilson coefficients $K_2^{L,R}$
in the model under consideration. In the SM extended with right-handed
neutrinos, the radiative process $\mu \to e \gamma$ is induced by
1-loop diagrams with $W$-bosons and neutrinos running in the loop, as
shown in Fig. \ref{fig:nuloop}. One finds~\cite{Ma:1980gm}
\begin{align}
K_2^L = & \frac{G_F}{2 \sqrt{2} \pi^2} m_\mu \sum_i U_{i\mu} U^\ast_{i e}  
 \, (F_1 + F_2)\\
K_2^R = & \frac{G_F}{2 \sqrt{2} \pi^2} m_e \sum_i U_{i\mu} U^\ast_{i e}  
 \, (F_1 - F_2)
\end{align}
Here, $U_{ij}$ denote the entries of the so-called PMNS matrix and
$F_1$ and $F_2$ are loop functions. Approximately, one has $F_1 \simeq
-\frac{1}{4}\left(\frac{m_\nu}{m_W}\right)^2$ and $F_2 \simeq 0$.
Finally, we are just left with the final step, the computation of the
observable. The relation between the Wilson coefficients $K_2^{L,R}$
and BR($\mu \to e \gamma$) is found to be~\cite{Hisano:1995cp}
\begin{equation}
\text{BR}\left( \mu \to e \gamma \right) =
\frac{\alpha m_{\mu}^5}{4 \Gamma_\mu} \left( |K_2^L|^2 + |K_2^R|^2 \right) \, ,
\end{equation}                 
where $\Gamma_\mu$ is the muon total decay width. This general
expression holds for all models. With this final step, the computation
concludes.

As we have seen, the main task to get a result for a flavor observable
in a new model is to derive analytical expressions for the relevant
Wilson coefficients. This is the most complicated and model dependent
part of the calculation, and it might be a very demanding task in some
cases. First, all masses and vertices involved in the computation must
be known analytically. Furthermore, in some models the number of
Feynman diagrams that contribute to a specific Wilson coefficient
could be very high, thus leading to lengthy (and prone to error)
computations, and renormalization group equations (RGEs) are sometimes
required to evaluate the Wilson coefficients at a given energy
scale. Finally, a proper numerical evaluation of the results is
usually welcome, as some numerical difficulties may appear in
particular limits of complicated loop functions. In order to solve
these practical problems we present \FlavorKit, a tool that allows for
a fully automatized calculation of a wide range of flavor observables
for several classes of models.

\section{FlavorKit}
\label{sec:flavorkit}

In this Section we give a general overview of the \FlavorKit tool. For
more details we refer to the manual \cite{Porod:2014xia}. The code is
public and can be downloaded from the website {\tt
  http://sarah.hepforge.org/FlavorKit.html}, where additional
information and regular updates are published.

\begin{figure}[tb]
\begin{center}
\includegraphics[width=0.9\linewidth]{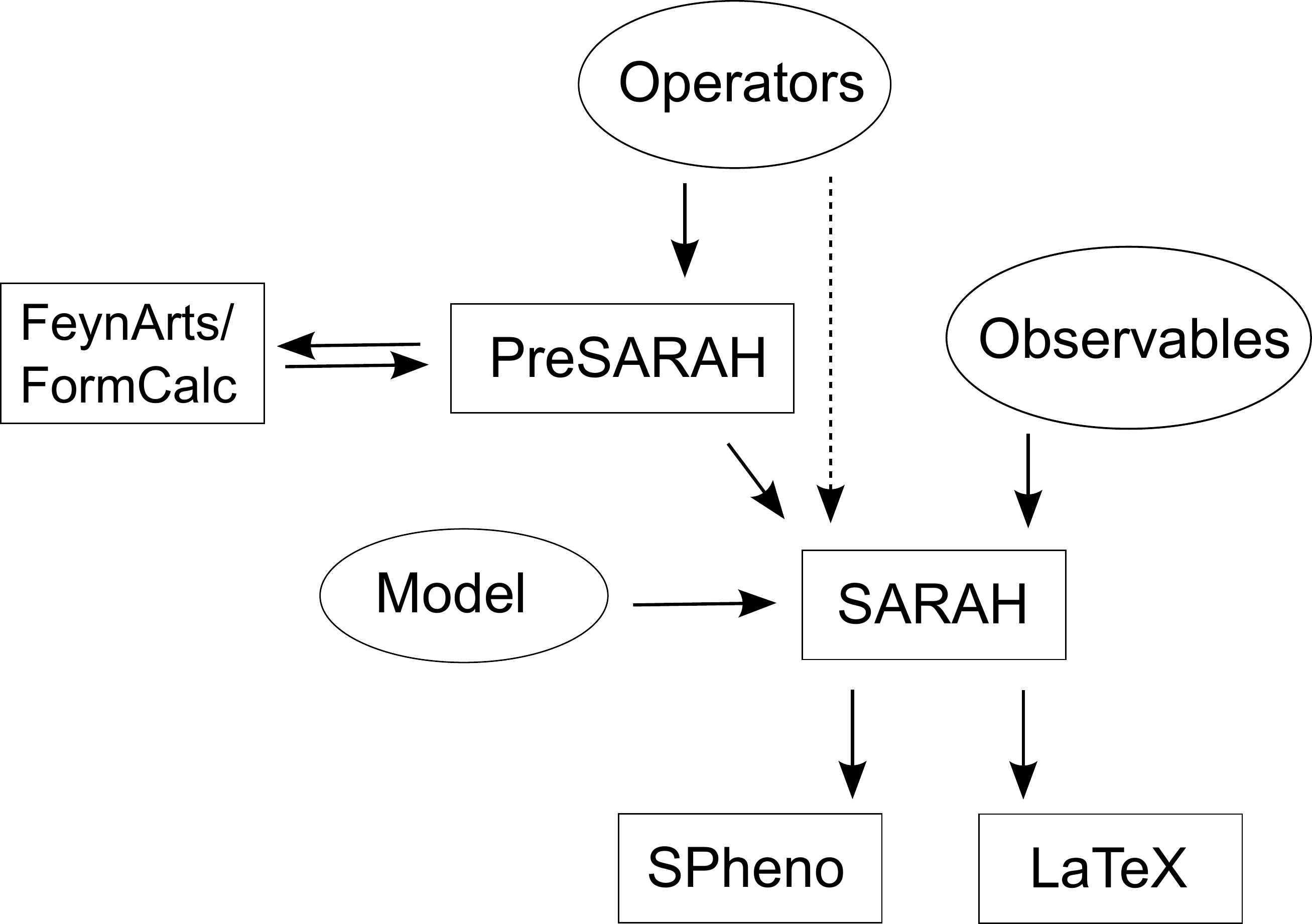}
\end{center}
\caption{Schematic description of \FlavorKit: the user can define new
  operators in \PreSARAH, which then calculates generic expressions
  for the Wilson coefficients using \FeynArts and \FormCalc and
  creates the necessary input files for \SARAH. In addition, \Fortran
  code is provided to relate the Wilson coefficients to specific
  flavor observables. This information is finally used by \SARAH to
  generate \SPheno code for the numerical calculation of the
  observables.}
\label{fig:setup}
\end{figure}

\FlavorKit is a combination of several public codes. The main
ingredients are \SARAH \cite{Staub:2013tta} and \SPheno
\cite{Porod:2003um,Porod:2011nf}. The former is a \Mathematica package
for building and analyzing BSM models. \SARAH can compute all
vertices, mass matrices, tadpoles equations, 1-loop corrections for
tadpoles and self-energies, as well as 2-loop RGEs in a given
model. This analytical information can then be transferred to \SPheno,
a \Fortran code for the numerical evaluation of the particle spectrum
as well as several observables. Furthermore, \FlavorKit also makes use
of \FeynArts/\FormCalc \cite{Hahn:2005vh,Nejad:2013ina} to compute
generic expressions for the required Wilson coefficients at the tree-
and 1-loop levels.

As previously discussed, the critical step in the computation of a
flavor observable is the diagrammatic computation of the Wilson
coefficients of the involved operators. For this purpose, we have
created a new \Mathematica package called \PreSARAH. This package
uses \FeynArts and \FormCalc to calculate generic 1-loop amplitudes,
to extract the coefficients of the demanded operators, to translate
them into the syntaxis needed for \SARAH and to write the necessary
wrapper code. Currently, \PreSARAH works for any 4-fermion or
2-fermion-1-boson operators and will be extended in the future to
include other kinds of operators. The current version contains a long
list of fully implemented operators, ready to be used in the
evaluation of flavor observables. The results obtained with
\PreSARAH are then interpreted by \SARAH, which adapts the generic
expressions to the specific details of the model chosen by the user
and uses snippets of \Fortran code to calculate flavor observables
from the resulting Wilson coefficients. Finally, \SARAH can be used to
obtain analytical output in \LaTeX\ format or to create \Fortran
modules for \SPheno, in case the user is interested in performing
numerical studies. The approach is described in Fig. \ref{fig:setup},
which shows a schematic description of \FlavorKit.

\FlavorKit can be used in two ways:

\begin{itemize}

\item {\bf Basic usage:} This is the approach to be followed by the
  user who does not need any operator nor observable beyond what is
  already implemented in \FlavorKit. In this case, \FlavorKit reduces
  to the standard \SARAH package. 

\item {\bf Advanced usage:} This is the approach to be followed by the
  user who needs an operator or an observable not included in
  \FlavorKit. The user can define his own operators and get analytical
  results for their coefficients using \PreSARAH. The resulting
  output can be passed to \SARAH and then continue with the basic
  usage.  In case the user is interested in a new observable, not
  implemented in \FlavorKit, he can also add it easily. The only
  requirement is the addition of a few lines of \Fortran code with the
  relation between the new observable and the Wilson coefficients of
  the operators that contribute (these can be implemented by default
  or added by the user). Then one can continue with the basic usage.

\end{itemize}

Let us give some more details about these two ways to use \FlavorKit.

\subsection{Basic usage}
\label{subsec:basic}

\begin{table}[tb]
\centering
\begin{tabular}{c c}
\hline
{\bf Lepton flavor} & {\bf Quark flavor} \\
\hline
$\ell_\alpha \to \ell_\beta \gamma$ & $B_{s,d}^0 \to \ell^+ \ell^-$ \\
$\ell_\alpha \to 3 \, \ell_\beta$ & $\bar B \to X_s \gamma$ \\
$\ell_\alpha \to \ell_\beta \ell_\gamma \ell_\gamma$ & $\bar B \to X_s \ell^+ \ell^-$ \\
$\mu-e$ conversion in nuclei & $\bar B \to X_{d,s} \nu \bar \nu$ \\
$\tau \to P \, \ell$ & $B \to K \ell^+ \ell^-$ \\
$h \to \ell_\alpha \ell_\beta$ & $K \to \pi \nu \bar \nu$ \\
$Z \to \ell_\alpha \ell_\beta$ & $\Delta M_{B_{s,d}}$ \\
 & $\Delta M_{K}$ and $\varepsilon_K$ \\
 & $P \to \ell \nu$ \\
\hline
\end{tabular}
\caption{List of flavor violating processes and observables which have
  been already implemented in \FlavorKit (version {\tt 1.0.1}). To the
  left, observables related to lepton flavor, whereas to the right
  observables associated to quark flavor.}
\label{tab:observables}
\end{table}

In the basic usage of \FlavorKit the user simply uses \SARAH. More
information about installation and usage of \SARAH can be found in the
manual \cite{Staub:2013tta}. For the list of observables already
implemented in \FlavorKit see Table \ref{tab:observables}.

The most recent versions of \SARAH (after version {\tt 4.2.0}) include
\FlavorKit routines that allow one to obtain \LaTeX\ files with the
analytical expressions for the Wilson coefficients. These are given
individually for each Feynman diagram contributing to the
coefficients, what helps identifying the individual weigths of the
different physical contributions. The resulting \LaTeX\ files are
saved in the folder
\begin{center}
{\tt  [\$SARAH]/Output/[\$MODEL]/EWSB/TeX/FlavorKit/ }
\end{center}
Finally, in case the user is interested in the numerical evaluation of
the flavor observables, a \SPheno module can be created. This follows
again the usual steps in \SARAH. The generated \Fortran code can be
used to run fast and reliable numerical studies of the parameter space
of the model.

\subsection{Advanced usage}
\label{subsec:advanced}

In some cases the user might be interested in new contributions to a
flavor observables induced by operators not considered in the
\FlavorKit implementation. Similarly, the user might also be
interested in new observables, not included in the current version of
the tool. In both cases, \FlavorKit can be easily extended to satisfy
all the user's requirements.

In order to introduce new observables in \FlavorKit, the user must
create two new files: a steering file with the extension {\tt .m}, and
a \Fortran body with the extension {\tt .f90}. These have to be placed
in the folder
\begin{center}
 {\tt [\$SARAH]/FlavorKit/[\$Type]/Processes/}
\end{center}
where {\tt [\$Type]} is either {\tt LFV} for lepton flavor violating
or {\tt QFV} for quark flavor violating observables. The steering file
provides basic information about the process. For instance, the
steering file to calculate the rate for $\ell_\alpha \to \ell_\beta
\gamma$ reads
\begin{lstlisting}
NameProcess = "LLpGamma";
NameObservables =
     {{muEgamma, 701, "BR(mu->e gamma)"}, 
      {tauEgamma, 702, "BR(tau->e gamma)"}, 
      {tauMuGamma, 703, "BR(tau->mu gamma)"}};
NeededOperators = {K2L, K2R};
Body = "LLpGamma.f90"; 
\end{lstlisting}
The three observables will be saved in the variables {\tt muEgamma},
{\tt tauEgamma} and {\tt tauMuGamma} and will show up in the output
file written by \SPheno in the block {\tt FlavorKitLFV}, with numbers
$701$ to $703$. Moreover, the steering file also refers to the body
file, in this case {\tt LLpGamma.f90}. This reads
\begin{lstlisting}[language=Fortran]
Real(dp) :: width
Integer :: i1, gt1, gt2

Do i1=1,3 

If (i1.eq.1) Then        ! mu -> e gamma
  gt1 = 2
  gt2 = 1
Elseif (i1.eq.2) Then    ! tau -> e gamma
  gt1 = 3
  gt2 = 1
Else                     ! tau -> mu gamma
  gt1 = 3
  gt2 = 2
End if

width = 0.25_dp*mf_l(gt1)**5*(Abs(K2L(gt1,gt2))**2 &
           & +Abs(K2R(gt1,gt2))**2)*Alpha

If (i1.eq.1) Then
  muEgamma = width/(width+GammaMu)
Elseif (i1.eq.2) Then 
  tauEgamma = width/(width+GammaTau)
Else
  tauMuGamma = width/(width+GammaTau)
End if

End do
\end{lstlisting}
Here {\tt Real(dp)} is the \SPheno internal definition of double
precision variables. If necessary, the user can also use complex
numbers with {\tt Complex(dp)}. This file contains standard \Fortran
{\tt 90} code relating the observables (previously declared in the
steering file) with the Wilson coefficients computed by \FlavorKit
($K_2^L$ and $K_2^R$ in this example). The relation can make use of SM
parameters, such as the fine structure constant, $\alpha$ ({\tt
  Alpha}), or the mass of the leptons, $m_\ell$ ({\tt
  mf\_l(gt1)}). These are internally defined in \FlavorKit, but can
also be changed by the user.

With this feature, the user can easily add new observables or modify
the implementation of the ones already present in the tool (see Table
\ref{tab:observables} for the list of observables already implemented
in \FlavorKit). For example, the user can change the values of some
parameters (like hadronic form factors or decay constants) and adapt
the calculation to his own needs. All it takes is to write a few lines
of \Fortran code with the definition of the observable in terms of the
Wilson coefficients that give a contribution.

The user can also implement new operators and obtain analytical
expressions for their Wilson coefficients. In this case, he will need
to use \PreSARAH, which provides generic expressions for the
coefficients, later to be adapted to specific models with \SARAH. In
order to do so, the user must extend the content of the folder
\begin{center}
 {\tt [\$SARAH]/FlavorKit/[\$Type]/Operators/}
\end{center}
where {\tt [\$Type]} is again either {\tt LFV} for operators leading
to lepton flavor violating processes or {\tt QFV} for operators that
induce quark flavor violation. The extension only involves one input
file. This file must contain the definition of the operator and some
basic details about its computation. For instance, the \PreSARAH input
file to calculate the Wilson coefficients of the $(\bar{\ell}\Gamma
\ell)(\bar{d} \Gamma' d)$ operators reads
\begin{lstlisting}
NameProcess="2L2d";
ConsideredProcess = "4Fermion";
ExternalFields={{ChargedLepton,bar[ChargedLepton],
                   DownQuark,bar[DownQuark]}};
                   
FermionOrderExternal={2,1,4,3};
NeglectMasses={1,2,3,4}; 
                   

AllOperators={
   (* scalar operators*)
   {OllddSLL,Op[7].Op[7]},
   {OllddSRR,Op[6].Op[6]},
   {OllddSRL,Op[6].Op[7]},
   {OllddSLR,Op[7].Op[6]},

   (* vector operators*)
   {OllddVRR,Op[7,Lor[1]].Op[7,Lor[1]]},
   {OllddVLL,Op[6,Lor[1]].Op[6,Lor[1]]},
   {OllddVRL,Op[7,Lor[1]].Op[6,Lor[1]]},
   {OllddVLR,Op[6,Lor[1]].Op[7,Lor[1]]},

   (* tensor operators*)
   {OllddTLL,Op[-7,Lor[1],Lor[2]].Op[-7,Lor[1],Lor[2]]},
   {OllddTLR,Op[-7,Lor[1],Lor[2]].Op[-6,Lor[1],Lor[2]]},
   {OllddTRL,Op[-6,Lor[1],Lor[2]].Op[-7,Lor[1],Lor[2]]},
   {OllddTRR,Op[-6,Lor[1],Lor[2]].Op[-6,Lor[1],Lor[2]]}
};

CombinationGenerations = {{2,1,1,1}, {3,1,1,1}, {3,2,1,1},
                          {2,1,2,2}, {3,1,2,2}, {3,2,2,2}};

Filters = {};                          
\end{lstlisting}
We neglect all external masses in the operators (using {\tt
  NeglectMasses=\{1,2,3,4\}}), and the different coefficients of the
scalar operators $(\bar{\ell}P_X \ell)(\bar{d} P_Y d)$ are denoted as
{\tt OllddSXY}, the ones for the vector operators $(\bar{\ell} P_X
\gamma_\mu \ell)(\bar{d} P_Y \gamma^\mu d)$ as {\tt OllddVYX} and the
ones for the tensor operators $(\bar{\ell} P_X \sigma_{\mu \nu}
\ell)(\bar{d} \sigma^{\mu \nu} P_Y d)$ as {\tt OllddTYX}, with
$X,Y=L,R$. To make the definitions of the operators more readable, we
have simplified the syntaxis of the {\tt DiracChain} object in {\tt
  FeynArts}/{\tt FormCalc}. \PreSARAH puts everything with the head
{\tt Op} into a Dirac chain, using the defined fermion order. For
4-fermion operators the combination of both operators is written as
dot product. For example, {\tt Op[6].Op[6]} is internally translated
into
\begin{lstlisting}
DiracChain[Spinor[k[1],MassEx1,-1],6,Spinor[k[2],MassEx2,1]]*
DiracChain[Spinor[k[3],MassEx3,-1],6,Spinor[k[4],MassEx4,1]]
\end{lstlisting}
For more details see the \FlavorKit manual
\cite{Porod:2014xia}. Finally, we explicitly choose not to calculate
all possible combinations of external states, but only some specific
cases of interest: $\mu e d d$, $\tau e d d$, $\tau \mu d d$, $\mu e s
s$, $\tau e s s$, $\tau \mu s s$.

These two possibilities, extending the list of observables and
extending the list of operators, makes \FlavorKit an adaptable code,
that can be used for the calculation of observables not considered by
the authors of the tool~\footnote{\FlavorKit was used in
  \cite{Abada:2014kba} to perform the first complete study of lepton
  flavor violation in models with a low-scale seesaw. This paper
  represents a perfect example of what \FlavorKit can offer.}.

Before concluding, let us briefly comment on the validation of
\FlavorKit. This involves three steps:
\begin{enumerate}
 \item {\bf Agreement with SM results}: we have explicitly checked
   that \FlavorKit reproduces correctly the SM predictions for the
   flavor observables in the literature.
 \item {\bf Independence of scale in loop function}: the loop
   integrals for 2- and 3-point functions ($B_i, C_i$) depend on the
   renormalization scale $Q$. However, we checked numerically that the
   sum of all diagrams is indeed independent of the choice of $Q$. In
   some cases the check was also done analytically, confirming the
   consistency of our results.
 \item {\bf Comparison with other tools}: we also did a detailed
   numerical comparison with other flavor tools using the \SPheno code
   produced by \SARAH for the Minimal Supersymmetric Standard Model
   (MSSM). We find that \FlavorKit agrees with the codes specialized
   for the MSSM to the same level as those codes agree among each
   other. The small numerical differences are caused by the treatment
   of the resummation of the bottom Yukawa couplings, the different
   order at which SM and supersymmetric contributions are implemented,
   the different handling of the Weinberg angle, and the different
   level at which the RGE running is taken into account by the tools.
\end{enumerate}
In conclusion, the validation of \FlavorKit is successful. The results
are in good agreement with those obtained with other flavor tools and
the required consistency checks are passed.

\section{Summary}
\label{sec:summary}

FlavorKit is a combination of computer tools that allows the user to
get predictions for his favourite flavor observables in the model of
his choice. It combines the analytical power of \SARAH with the
numerical routines of \SPheno, leading to the perfect tool for
phenomenological studies. Furthermore, it is easily extendable: new
observables and operators can be added (thanks to \FeynArts and
\FormCalc). This results in a user friendly computer tool, ready for
the study of flavor physics beyond the Standard Model.

\section*{Acknowledgements}

The author is grateful to Werner Porod and Florian Staub for their
collaboration in the development of \FlavorKit.

\nocite{*}
\bibliographystyle{elsarticle-num}
\bibliography{refs}

\end{document}